\documentclass[twocolumn,showpacs,preprint2,amsmath,amssymb,aps]{revtex4}
\usepackage{graphicx}
\usepackage{dcolumn}
\usepackage{epsfig}
\usepackage{bm}
\usepackage{textcomp}
\usepackage{hyperref}

\begin{document}

\title{Quasielastic contribution to antineutrino-nucleus scattering}

\author{V.~Pandey\footnote{Vishvas.Pandey@UGent.be}, N.~Jachowicz\footnote{Natalie.Jachowicz@UGent.be}, 
J.~Ryckebusch, T.~Van Cuyck, W.~Cosyn}

\affiliation{Department of Physics and Astronomy,\\ Ghent University, \\Proeftuinstraat 86, \\ B-9000 Gent, Belgium.\\}


\begin{abstract}
We report on a calculation of cross sections for charged-current
quasielastic antineutrino scattering off $^{12}$C in the energy
range of interest for the MiniBooNE experiment. We adopt the impulse
approximation (IA) and use the nonrelativistic continuum random phase
approximation (CRPA) to model the nuclear dynamics. An effective
nucleon-nucleon interaction of the Skyrme type is used. We compare our
results with the recent MiniBooNE antineutrino cross-section data and
confront them with alternate calculations. The CRPA predictions
reproduce the gross features of the shape of the measured double-differential cross sections. 
The CRPA cross sections are typically larger than those of other
reported IA calculations but tend to underestimate the magnitude of the
MiniBooNE data.
We observe that an enhancement of the nucleon axial
mass in CRPA calculations is an
effective way of improving on the description of the shape and
magnitude of the double-differential cross sections.  The rescaling of
$M_{A}$ is illustrated to affect the shape of the double-differential cross
sections differently than multinucleon effects beyond the IA.
\end{abstract}


\pacs{25.30.Pt, 13.15.+g, 24.10.Cn, 21.60.Jz}

\maketitle


\section{Introduction}
Recent times have been marked by a substantial increase in the amount of data for
(anti)neutrino-nucleus interactions at intermediate energies.  Recently, the
MiniBooNE collaboration has published their first charged-current
quasielastic (CCQE) antineutrino-nucleus scattering cross-section
measurements \cite{minibooneantinu}. The underlying reaction process
of CCQE with antineutrino beams is $\bar{\nu}_{\mu}+p \rightarrow
\mu^{+}+n$ on bound protons.  Antineutrino-nucleus ($\bar{\nu} A$)
cross sections are less well measured than their neutrino
counterparts, mainly because of higher background contributions and
smaller statistics~\cite{revzeller}.  MiniBooNE has also published
cross sections for CCQE neutrino ($\nu_{\mu}+n \rightarrow \mu^{-}+p$)
~\cite{miniboonenucc} and neutral-current quasielastic (NCQE)
neutrino ($\nu_{\mu}+N \rightarrow \nu_{\mu}+N$) ~\cite{miniboonenunc}
processes. Several other collaborations have been contributing to the
increase of the neutrino-nucleus cross section database in recent
times. For example, T2K has released inclusive CC neutrino ~\cite{t2knu} data, 
whereas MINER$\nu$A presented CC neutrino ~\cite{minervnu} and
antineutrino ~\cite{minervanu} cross section results.

The modeling of $\nu A$ and $\bar{\nu} A$ scattering data poses some
real challenges.  In contrast to electron-nucleus scattering data for
which the initial electron energy is exactly known, the $\nu A$ and
$\bar{\nu} A$ data are $\nu$ ($\bar{\nu}$)-flux integrated
~\cite{revmorfin}. Despite the enormous improvements in the
experimental and theoretical understanding of (anti)neutrino-nucleus
interactions in the few GeV region, the current experimental precision
is of the order of 20 \textendash~30\% and the underlying processes on bound nucleons 
are not fully understood ~\cite{revmorfin,revzeller,lalakulich2012,benhar2010}.
Theoretical predictions for MiniBooNE's $\bar{\nu}_{\mu} + ^{12}$C
measurements are reported in Refs.~\cite{nieves,martini,ivanov,Amaro:2011aa,Meucci:2012yq}. 
References~\cite{nieves,martini} adopt a rather basic nuclear-structure model which cannot be expected to capture the complexity of the 
nuclear dynamics at low nuclear excitation energies. Reference~\cite{ivanov} starts from a relativistic mean-field model for the 
bound and scattering states. The approach in Ref.~\cite{Amaro:2011aa} is based on superscaling Approximation and Ref.~\cite{Meucci:2012yq}
adopts a relativistic Green's function model.
Reference~\cite{martini}
computes nuclear response functions with a local Fermi gas model in the
random phase approximation (RPA) and incorporates multinucleon effects
exclusively in the spin-isospin channels. Reference~\cite{nieves} starts
from a local Fermi gas description of the nucleus and includes
RPA correlations and multinucleon effects.  Both calculations for the 
$\bar{\nu}_{\mu} + ^{12}$C responses stress
the importance of multinucleon mechanisms at MiniBooNE kinematics, and
adopt a value for the axial mass ($M_A~\approx$ 1 GeV) in a dipole parametrization
of the axial form factor, which is consistent with the one used to
model the QE contribution to ${\nu}_{\mu} + ^{12}$C
~\cite{martini5,martini2,martini3,nieves2}. The multinucleon mechanisms account
for mechanisms in the $W$-nucleus coupling beyond the impulse
approximation (IA). In the IA, the $W$-nucleus coupling is
approximated as a sum of one-body $W$-nucleon couplings. Effects
beyond the IA introduce some uncertainties in the calculations,
particularly for finite nuclei as a consistent treatment of the
multinucleon electroweak currents is extremely challenging. According
to a recent study of neutrino scattering off the deuteron the effect
of two-body currents (excluding pion production channels) is smaller
than 10\% \cite{shen2012}.

\begin{figure*}
\includegraphics[bb=112 65 350 273, width=0.32\textwidth]{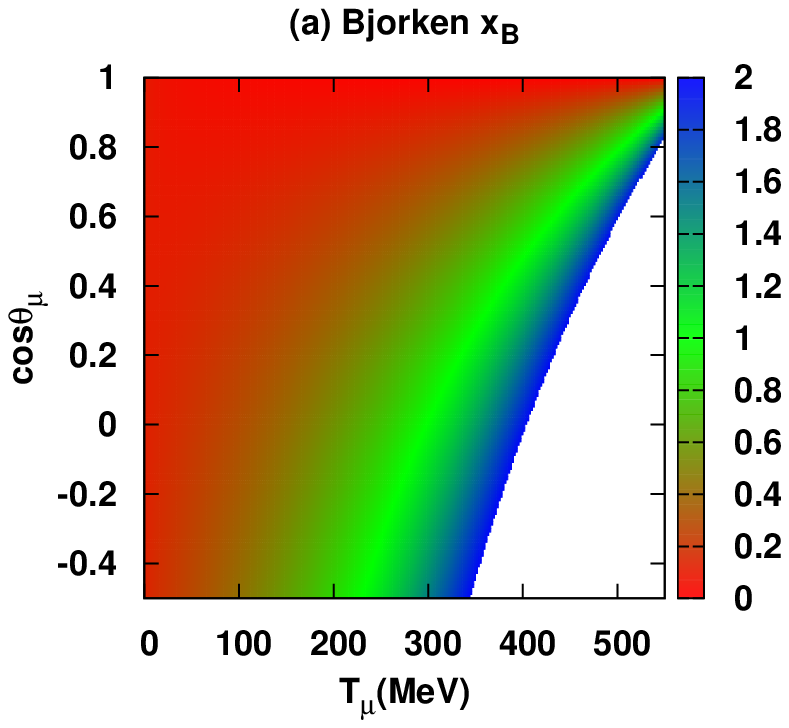}
\includegraphics[bb=112 65 350 273, width=0.32\textwidth]{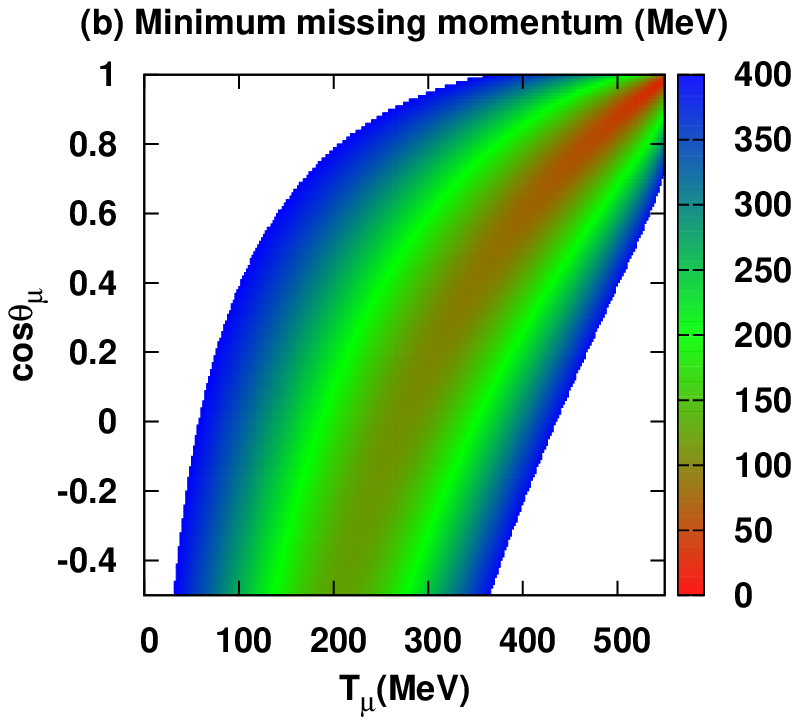}
\includegraphics[bb=112 65 350 273, width=0.32\textwidth]{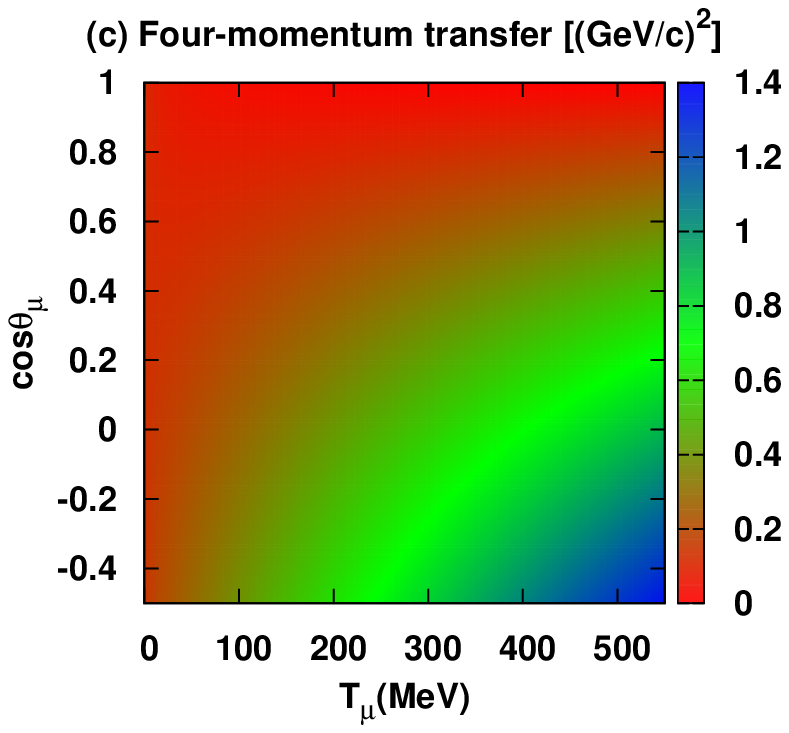}
\caption{(Color online) The kinematic variables (a) Bjorken $x_B$,
  (b) minimum $p_{mis}$, and (c) $Q^2$ as a function of
  $T_{\mu}$ and $\cos\theta_{\mu}$ at
  $E_{\bar{\nu}_{\mu}}$=700~MeV. White regions correspond with values
  of the variables out of the specified ranges.}
\label{fig1}
\end{figure*}

In this work, we adopt the IA for modeling the electroweak-nucleus
coupling and use a more sophisticated model for describing the
structure of the initial and final nuclei. In our approach to
investigate MiniBooNE's CCQE $\bar{\nu}_{\mu} + ^{12}$C results, we
model the nuclear dynamics starting from the mean field (MF)
description and introduce long-range correlations by means of a
nonrelativistic continuum RPA (CRPA) framework. Thereby, we use
Green's functions (or propagators) to solve the CRPA equations and an
effective Skyrme nucleon-nucleon residual interaction. The model takes
into account one-particle one-hole (1p-1h) excitations out of a
correlated nuclear ground state. In the CRPA, the effects of final-state
interactions of the ejected nucleons with the residual nucleus
are implemented. Thereby, one accounts for both distortions on the ejected
nucleon waves and rescatterings with the residual A-1 nucleons. For
example, rescattering effects $\bar{\nu}_{\mu}+p+(A-1) \rightarrow
\mu^{+}+n +(A-1) \rightarrow \mu^{+}+n^{\prime} +(A-1) {^\prime}$ are included. In CRPA the strength of the rescatterings is
regulated with the residual nucleon-nucleon force.
In the results section we focus on the influence of RPA correlations on the 
computed antineutrino responses for the MiniBooNE kinematics. The CRPA 
formalism does not contain relativistic corrections in its description of the nuclear
dynamics. In Refs.~\cite{amaro2005, amaro2007, martini2} one proposes to correct 
the energy transfer $\omega$ to account for relativistic effects in non-relativistic Fermi-gas
calculations. These methods, however, cannot be readily applied to the CRPA framework, 
as the computed response scales with the asymptotic nucleon kinetic energies in a complicated fashion. It is
  worth mentioning that MiniBooNE's antineutrino flux distribution is
  shifted to lower energies compared to the neutrino one.  Therefore,
  it can be anticipated that the $\overline{\nu}_{\mu}+^{12}$C responses are
  subject to smaller relativistic corrections than the ${\nu}_{\mu}+^{12}$C
  ones.

The paper is organized as follows. In Sec.~\ref{frmlsm}, we briefly
describe the CRPA framework of our cross-section calculations. In
Sec.~\ref{res}, we present numerical results of $\bar{\nu}_{\mu} +
^{12}\text{C}$ cross sections and compare them with the MiniBooNE data
and with other theoretical models. The conclusions are given in Sec.~\ref{conc}.

\begin{figure}
\includegraphics[width=0.8\columnwidth]{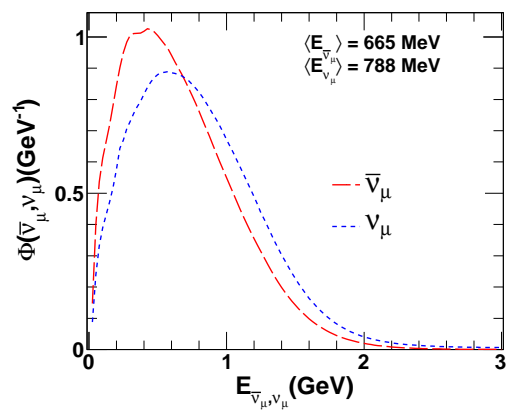}
\caption{(Color online) The MiniBooNE antineutrino and neutrino flux
  \cite{minibooneantinu,miniboonenucc} normalized to $1$.}
\label{fig2}
\end{figure}


\section{Formalism}\label{frmlsm}

In this work, we focus on the inclusive CCQE antineutrino nuclear reaction
\begin{equation}
 \bar{\nu}_{\mu}+\, ^{12}\text{C} \rightarrow \mu^{+}+X~,
\end{equation}
with no pion in the final state, a process which is referred to as
QE-like \cite{martini5,martini2,nieves3}. We obtain nuclear responses
with the CRPA method, which is described in details in
Refs.~\cite{neut,char}. This formalism has been successfully used in
the description of exclusive photo-induced and electro-induced QE
processes ~\cite{jan1,jan2} and in inclusive neutrino scattering at
supernova energies~\cite{neut,char,natproc,nuintproc}. Here, the CRPA
method is applied to antineutrino-nucleus interactions at intermediate
energies. The CRPA framework includes all single-nucleon knockout
channels and is therefore well suited to compute the quasielastic
contribution to the inclusive (anti)neutrino-nucleus responses. The
CRPA framework is not suited to compute the contributions from
alternate reaction mechanisms such as multinucleon knockout.

\begin{figure}
\includegraphics[width=0.8\columnwidth]{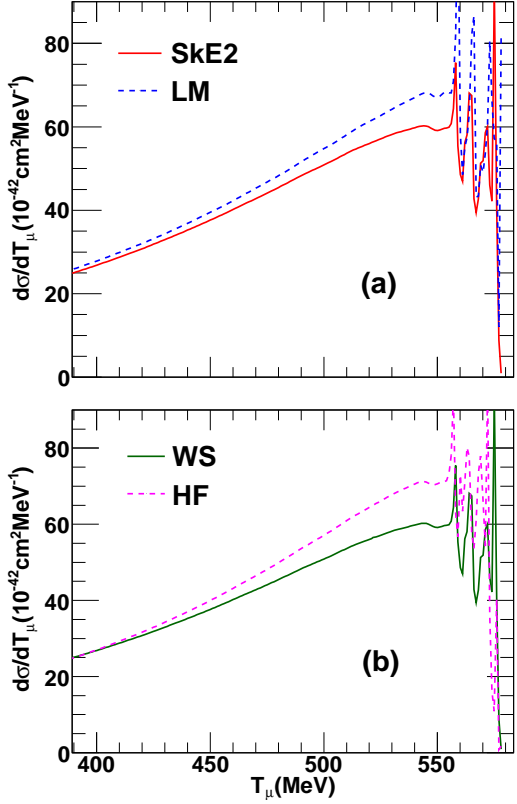}
\caption{(Color online) The $T_{\mu}$ dependence of the QE
  antineutrino-$^{12}$C CRPA cross sections for $E_{\bar{\nu}_{\mu}}=$
  700~MeV. (a) Curves obtained with the SkE2 and
  Landau-Migdal (LM) residual interaction with WS as single-particle wave functions. (b)
  Curves obtained with the WS and HF single-particle wave functions with SkE2 as residual interaction. }
\label{fig9}
\end{figure}

We summarize the basis ingredients of the model. An effective Skyrme
two-body interaction (more specifically, the SkE2 parametrization
\cite{jan1}) is used to construct a mean-field (MF) potential. The
bound and continuum single-nucleon wave functions can be obtained as
the solutions to the corresponding Schr\"{o}dinger equation.
The long-range correlations between the nucleons are introduced
through the RPA which describes an excited nuclear state with a
nucleon in the energy continuum of the MF potential as the coherent
superposition of particle-hole (ph$^{-1}$) and hole-particle
(hp$^{-1}$) excitations out of a correlated ground state, which has
0p-0h and 2p-2h components
\begin{equation}
 \arrowvert \Psi_{RPA}^{C} \rangle = \sum_{C^{'}} \left\{ X_{C,C^{'}} ~ \arrowvert ph^{-1} \rangle 
                                  -  ~Y_{C,C^{'}}~ \arrowvert hp^{-1} \rangle \right\}~.
\end{equation}
Here, $C^{'}$ stands for a combination of all quantum numbers of a hole and particle
state. Green's function theory allows one to treat the single-particle
energy continuum exactly \cite{jan1}. In computing the response of the
nucleus to an external electroweak probe a key quantity is the RPA
polarization propagator which can be obtained as a solution to the following
iterative equation~:
\begin{eqnarray}
 & & \Pi^{(RPA)} (x_1,x_2;E_x) =  \Pi^{(0)} (x_1,x_2;E_x) \nonumber \\
&   &  + \frac{1}{\hbar} \int dx dx' \Pi^{0} (x_1,x;E_x)  
\tilde{V}(x, x') \Pi^{(RPA)} (x',x_2;E_x) , \nonumber \\ &&
\end{eqnarray}
where $E_x$ is the excitation energy of the target nucleus and $x$ is
a shorthand notation for the combination of the spatial, spin and
isospin coordinates. Further, $\Pi^{(0)}$ denotes the MF contribution to
the polarization propagator and $\tilde{V}$ is the antisymmetrized
residual interaction. The MF responses can be computed by neglecting
the second term in the above equation. The second term accounts for
the multiple-scattering events after the initial electroweak
excitation of a nucleon from a bound into a continuum state. In the MF
approach, only direct nucleon knockout is included and the sole
implemented final-state interaction (FSI) effect is the distortion of the ejected-nucleon waves
in the real MF potential of the residual nucleons.

In terms of the experimentally measured quantities (outgoing muon
kinetic energy $T_{\mu}$ and cosine of the muon scattering angle
$\cos{\theta}_{\mu}$), the twofold differential cross section for CC
(anti)neutrino-nucleus scattering is given by~:
\begin {eqnarray}
&  & \left(\frac{d^2\sigma}{dT_{\mu} d\cos\theta_{\mu} } \right)_{\nu,\bar{\nu}}  =  G_{F}^{2} ~\cos ^{2} {\theta_c}~\left(\frac{2}{2J_i+1}\right)~
                                                                                   \varepsilon_{\mu}^{2} ~{\tilde{k}_{\mu}} \nonumber \\
&  & ~~~~~~~~~~~~~~~~~~\times~ F \left(Z',  \varepsilon_{\mu}\right) \left[ \sum_{J=0}^{\infty} \sigma_{CL}^{J} +
                                                                                     \sum_{J=1}^{\infty} \sigma_{T}^{J} \right] , \label{diffcs}
\end {eqnarray}
where $G_{F}$ is the weak interaction coupling constant and
$\theta_{c}$ is the Cabibbo angle. Further, $\tilde{k}_{\mu} = k_{\mu}/\varepsilon_{\mu}$
with $k_{\mu}$ ($\varepsilon_{\mu}$) is
the momentum (energy) of the final lepton. The Fermi function $F \left(Z',
\varepsilon_{\mu} \right)$, is introduced in order to take into account the
Coulomb interaction between the outgoing lepton and the residual
nucleus which has a proton number $Z'$. In order to compute the
differential cross sections we rely on a multipole expansion of the
weak transition operators and in Eq.~(\ref{diffcs}) the
$\sigma_{CL}^{J}$ and $\sigma_{T}^{J}$ are the Coulomb longitudinal
and the transverse contributions for a given multipolarity $J$~:
\begin{eqnarray}
 \sigma_{CL}^{J} & = & v^{\mathcal{M}} ~ |\langle J_f|| \widehat{\mathcal{M}}_J(|\vec{q}|)|| J_i \rangle|^2
                     + v^{\mathcal{L}} ~ |\langle J_f|| \widehat{\mathcal{L}}_J(|\vec{q}|)|| J_i \rangle|^2 \nonumber \\
                 &   & +~2 ~v^{\mathcal{M}\mathcal{L}} ~\mathcal{R}\left[\langle J_f|| \widehat{\mathcal{L}}_J(|\vec{q}|)|| J_i \rangle \langle J_f|| \widehat{\mathcal{M}}_J(|\vec{q}|)|| J_i \rangle^{\ast} \right], \nonumber \\
\end{eqnarray}
with
\begin{eqnarray}
 v^{\mathcal{M}} = \left[1+\tilde{k}_{\mu}\cos\theta_{\mu}\right], \nonumber
\end{eqnarray}

\begin{eqnarray}
 v^{\mathcal{L}} = \left[1+\tilde{k}_{\mu}\cos\theta_{\mu}- \frac{2 \varepsilon_i \varepsilon_{\mu}}{|\vec{q}|^2} \tilde{k}_{\mu}^2 \sin^{2}\theta_{\mu}\right], \nonumber
\end{eqnarray}
 
 \begin{eqnarray}
 v^{\mathcal{M}\mathcal{L}} = \left[\frac{\omega}{|\vec{q}|}(1+\tilde{k}_{\mu}\cos\theta_{\mu})+ \frac{m_{\mu}^{2}}{\varepsilon_{\mu}|\vec{q}|}\right], \nonumber
\end{eqnarray}
and 
\begin{eqnarray}
  \sigma_{T}^{J} & = & v^{T} \left[ |\langle J_f|| \widehat{\mathcal{J}}_J^{mag}(|\vec{q}|)|| J_i \rangle|^2
                      + |\langle J_f|| \widehat{\mathcal{J}}_J^{el}(|\vec{q}|)|| J_i \rangle|^2 \right] \nonumber \\
                   &   & \mp ~2 ~v^{TT} ~\mathcal{R}\left[\langle J_f|| \widehat{\mathcal{J}}_J^{mag}(|\vec{q}|)|| J_i \rangle \langle J_f|| \widehat{\mathcal{J}}_J^{el}(|\vec{q}|)|| J_i \rangle^{\ast} \right], \label{sigmaT} \nonumber \\
\end{eqnarray}
with
\begin{eqnarray}
 v^{T} = \left[1-\tilde{k}_{\mu}\cos\theta_{\mu}+\frac{\varepsilon_i \varepsilon_{\mu}}{|\vec{q}|^2} \tilde{k}_{\mu}^2 \sin^{2}\theta_{\mu}\right],  \nonumber
\end{eqnarray}

 \begin{eqnarray}
 v^{TT} = \left[\frac{\varepsilon_i+\varepsilon_{\mu}}{|\vec{q}|}(1-\tilde{k}_{\mu}\cos\theta_{\mu})-\frac{m_{\mu}^{2}}{\varepsilon_{\mu}|\vec{q}|}\right]. \nonumber
\end{eqnarray}

Here, $Q^{2} = - q ^ {\mu}q_{\mu} $, with $q ^{\mu} \left(\omega,
\vec{q}\right)$ the transferred four-momentum carried by the $W$
boson. $\varepsilon_i$ is energy of the incoming neutrino and $m_{\mu}$ is the 
mass of the final lepton. The $\widehat{\mathcal{M}}_J$, $\widehat{\mathcal{L}}_J$,
$\widehat{\mathcal{J}}_{J}^{el}$ and $\widehat{\mathcal{J}}_{J}^{mag}$
denote the Coulomb, longitudinal, transverse electric and transverse
magnetic transition operators as defined in
Refs.~\cite{neut, char}. The $|\vec{q}|$ is the magnitude of the
transferred three-momentum and $J_{i}$ ($J_{f}$) represents the total
angular momentum of the initial (final) state of the nucleus.  The difference
between the neutrino and antineutrino CC cross section stems from the
sign assigned to the interference term in Eq.~(\ref{sigmaT}): positive
for the neutrino and negative for the antineutrino beams.

\begin{figure*}[bth]
\includegraphics[width=\textwidth]{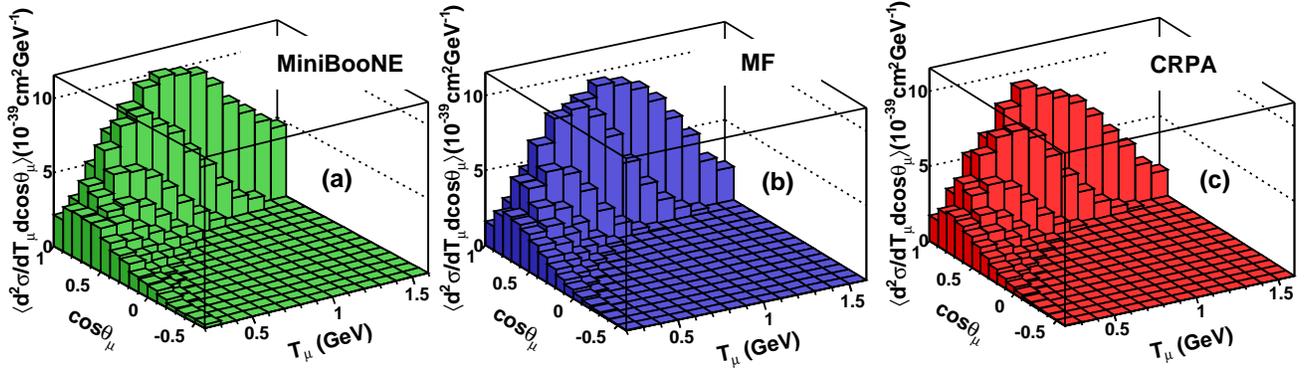}
\caption{(Color online) Double-differential cross section per target
  proton for $^{12}$C$(\bar{\nu}_{\mu}, \mu^{+})X$, as a function of
  $T_{\mu}$ and $\cos\theta_{\mu}$. The MiniBooNE data
  \cite{minibooneantinu} are plotted without the shape uncertainty and
  also excludes the 17.2\% normalization uncertainty. CRPA and MF
  calculations are folded with MiniBooNE $\bar{\nu}_{\mu}$
  flux.}
\label{fig3}
\end{figure*}

As mentioned, in this work we adopt the IA. Now we introduce a number
of variables which allow one to assess the validity of this
approximation for given kinematic settings.  The Bjorken $x_B$ scaling
variable is given by the invariant quantity
\begin{equation}
 x_B~=~\frac{A Q^{2}}{2p_{A}^{\mu}q_{\mu}}~,
\end{equation}

where $p_{A}^{\mu}$ is the momentum of the target nucleus. Figure ~\ref{fig1} displays 
$x_B$ as a function of the
experimentally measured quantities $T_{\mu}$ and $\cos{\theta}_{\mu}$
for $E_{\bar{\nu}_{\mu}} = 700$~MeV. 
As shown in Fig.~\ref{fig2}, MiniBooNE's $\bar{\nu}_{\mu}$
energy spectrum reaches its mean near 700~MeV. For $x_B \approx 1$, QE
single-nucleon knockout is expected to dominate and IA
calculations are expected to perform best. From Fig.~\ref{fig1} it
is clear that at very forward $\theta _{\mu}$ one expects the bulk of
the single-nucleon knockout strength at larger $T_{\mu}$.  With
increasing $\theta _{\mu}$  the QE single-nucleon knockout strength
will shift to lower $T_{\mu}$.  At kinematic conditions corresponding
with both low $T_{\mu}$ and forward muon scattering angles, one could
expect major contributions beyond the IA.

\begin{figure*}

\begin{minipage}{\dimexpr\linewidth-0.50cm\relax}
\rotatebox{90}{\mbox{\large \bm{$\textlangle d^{2}\sigma/dT_{\mu}dcos\theta_{\mu} \textrangle (10^{-42}cm^{2}MeV^{-1})$}}}
\vspace*{-8.3cm}\hspace*{16.8cm}
\end{minipage}%

\begin{minipage}{\dimexpr\linewidth-0.50cm\relax}
\includegraphics[width=0.94\textwidth]{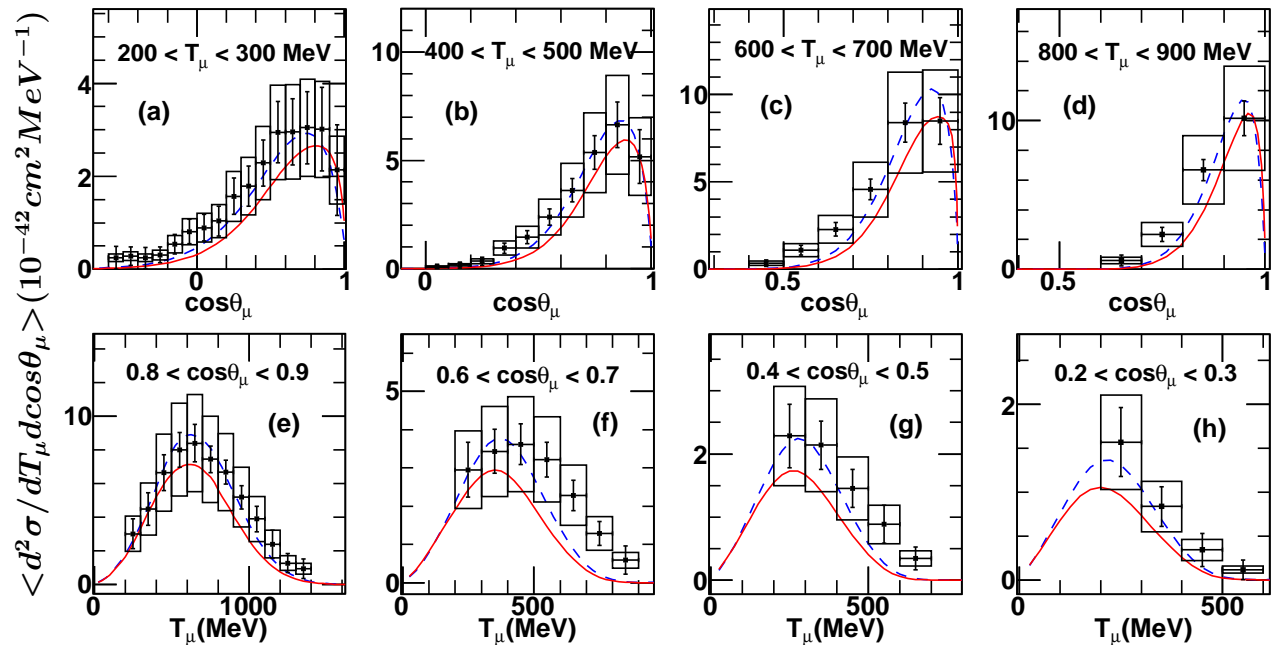}
\caption{(Color online) MiniBooNE flux-folded double-differential
  cross section per target proton for $^{12}$C$(\bar{\nu}_{\mu},
  \mu^{+})X$ plotted as a function $T_{\mu}$  for different ranges of
  $\cos\theta_{\mu}$ (bottom), as a function of $\cos\theta_{\mu}$ 
  for different $T_{\mu}$ values (top).  
  Solid curves are CRPA and dashed curves are MF
  calculations.  MiniBooNE data are filled squares, error bars
  represent the shape uncertainties and error boxes
  represent the 17.2\% normalization uncertainty.}
  \label{fig4}
\end{minipage}
\end{figure*}

The $x_B$ is a model-independent kinematic variable. We now introduce
a kinematic variable which is a highly relevant one for QE processes.
In direct single-nucleon knockout reactions, the momentum of the
initial bound nucleon (often referred as the missing momentum
$p_{mis}$) is the scaling variable \cite{Amaro2006}. Indeed, in the
plane-wave limit, the exclusive single-nucleon knockout cross sections
are directly proportional to the momentum distribution of the bound
nucleons in the target nucleus. Mean-field nucleons are characterized
by a momentum distribution which is Gaussian-like and extends over a
specific range ($ 0 \le p_{mis} \lesssim 250$~MeV) \cite{Maarten2012}.
Large missing momenta necessarily lead to small single-nucleon
knockout cross sections and/or substantial contributions from
competing multinucleon processes. Imposing a QE reaction process
($W^-+ p $ with $A-1$ spectators), energy and momentum conservation in the
laboratory frame can be expressed as
\begin{equation}
M_A + \omega = E_{A-1}^{\star}+ \sqrt{M_{n}^2 + p_n^2} , 
\hspace{0.5cm} \vec{p}_{mis} + \vec{q} = \vec{p}_n \; ,
\label{eq:conservationIA}
\end{equation} 
where $\vec{p}_n$ is the three-momentum of the ejected neutron in the
laboratory frame, $M_n$ is the neutron mass, and $M_{A}$ the mass of the target 
nucleus. The $E_{A-1}^{\star}$ is the total 
energy of the residual nucleus and includes contributions from recoil
and excitation energy
$E_{A-1}^{\star}=M_{A-1}+T_{A-1}+E_{exc}^{\star}$. The $p_{mis}$
depends on $\theta_{p_{n}q}$, the angle between $\vec{q}$ and
$\vec{p}_n$. For inclusive reactions as those considered in this
work, the relative importance of the quasielastic contribution can be
estimated with the aid of the minimum missing momentum: the minimum
value of $p_{mis}$ as $\theta_{p_{n}q}$ varies between 0$^{\circ}$ and
180$^{\circ}$.  In Fig.~\ref{fig1} we also display the minimum value
of the missing momentum, denoted as $p_{mis}^{min}$ for a given incoming neutrino energy and
$T_{A-1}+E_{exc}^{\star}=25$~MeV. As one moves along the $x_{B} \approx 1$
region, with increasing $\theta _{\mu}$ a
shift to larger $p_{mis}^{min}$ is observed and larger multinucleon
contributions can be expected \cite{stijnenjan}. The $\left( T_{\mu},
\cos \theta _{\mu} \right)$ kinematic settings with a minimum $p_{mis}
\gtrapprox$250~MeV are prone to multinucleon corrections beyond the
IA.  For the sake of completeness, we also show a contour plot of the $W$ boson's 
virtuality.  Kinematic regions with the lowest $Q^2$ exhibit the strongest sensitivity 
to collective nuclear structure mechanisms.

The wide range of values of ($x_{B}$, $p_{mis}^{min}$, $Q^{2}$) probed
in the MiniBooNE $\bar{\nu}_{\mu} + ^{12}$C experiment, presents real
challenges to the theoretical models. Accordingly, one can expect
 rather divergent views about the impact of various reaction mechanisms.


\section{Results}\label{res}

Various studies have attempted to bring the predictions of 
(anti)neutrino-nucleus models in accordance with experimental data.  
Several modifications of the IA-based models have been considered, 
including the enhancement of the axial mass $M_A$ and the introduction of  
multinucleon effects ~\cite{martini5,nieves2}.
These approaches have similar effects on neutrino scattering cross sections, 
bringing predictions closer to data. This impedes extraction 
of $M_A$ directly from the data and makes it difficult to use data to constrain 
the importance of multinucleon effects.
In the following, we seek to shed light on these issues by making an
analysis of QE cross sections and the relative importance of different contributions 
to neutrino and antineutrino scattering processes.  We will show that multinucleon 
contributions and an enhanced axial mass affect the shape of the cross section 
differently and alter neutrino and antineutrino cross sections in a different way.

\begin{figure}
\includegraphics[width=0.8\columnwidth]{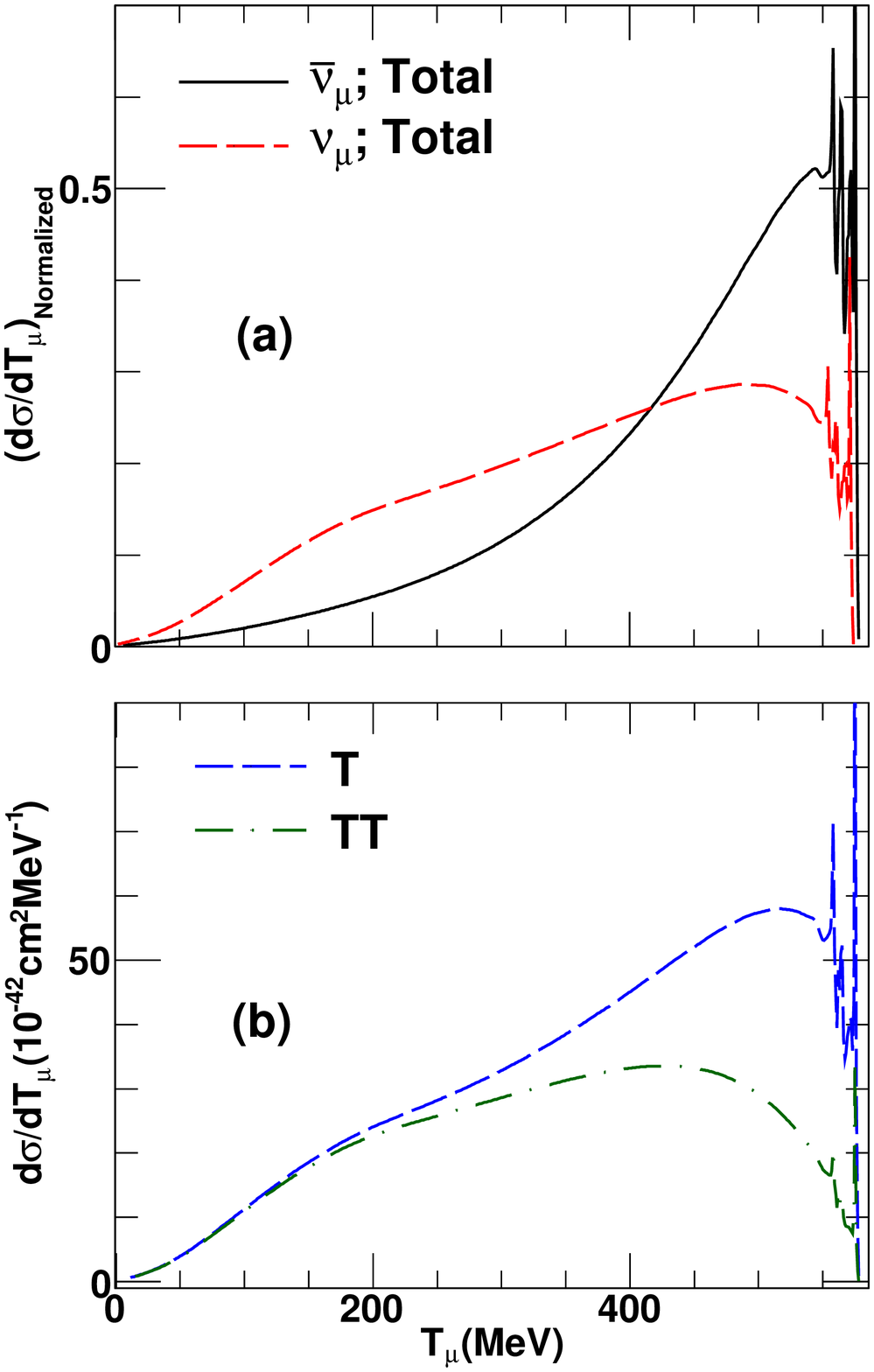}
\caption{(Color online) The $T_{\mu}$ distribution of the CRPA
  $^{12}$C$(\bar{\nu}_{\mu}, \mu^{+})$ and $^{12}$C$({\nu}_{\mu},
  \mu^{-})$ cross sections at a (anti)neutrino energy of 700~MeV.  
  (a) Total cross sections normalized to 1.  (b)
  Transverse contribution excluding the interference
  part (T) and the transverse interference (TT) contribution.}
\label{fig6}
\end{figure}

\begin{figure}
\includegraphics[width=0.8\columnwidth]{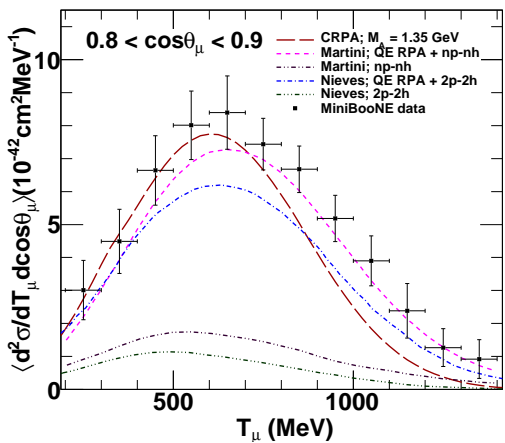}
\caption{(Color online) MiniBooNE flux-folded cross section per target
  proton for $^{12}$C$(\bar{\nu}_{\mu}, \mu^{+})X$ at $0.8 <
  \cos\theta_{\mu} < 0.9$. The CRPA predictions are compared with those of 
  Refs.~\cite{martini} and ~\cite{nieves}.}
\label{fig11} 
\end{figure}
\begin{figure}
\includegraphics[width=0.8\columnwidth]{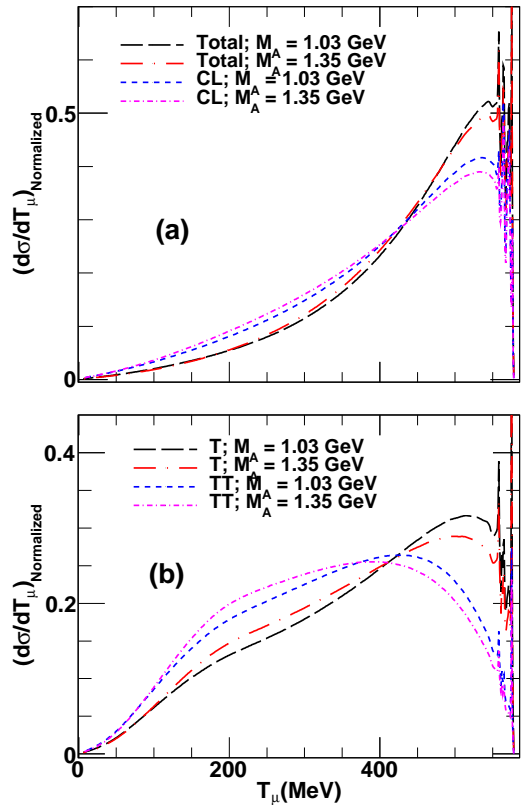}
\caption{(Color online) Normalized Coulomb longitudinal (CL), transverse term
  without interference (T) and transverse interference (TT) contributions to
  $^{12}$C$(\bar{\nu}_{\mu}, \mu^{+})X$ as a function of $T_{\mu}$ for two different values of
  $M_{A}$, at $E_{\bar{\nu}_{\mu}}$ = 700 MeV.}
\label{fig7}
\end{figure}

In order to test the robustness of calculations, we first investigate 
their sensitivity  to the nuclear physics input.
In computing the electroweak responses with the CRPA method, input is
required with regard to the residual nucleon-nucleon interactions, the
mean-field wave functions, and mean-field potential.  In
Fig.~\ref{fig9}, the sensitivity of the computed cross sections to the
nuclear-physics input is studied at $E_{\bar{\nu}_{\mu}}=$700~MeV. In
the top panel, we compare cross sections obtained with a Skyrme (SkE2) 
~\cite{jan2,skyrme2} and a Landau-Migdal parametrization ~\cite{landmig} 
for the residual effective nucleon-nucleon (NN) force.  The
sensitivity to the NN force is small for low outgoing muon energies but becomes
substantial at higher $T_{\mu}$, corresponding to lower nuclear excitation 
energies where differences amount to almost 15\%. This is expected
as it corresponds with a kinematic range most prone to nuclear collective
effects. At low $T_{\mu}$ the cross sections are
rather insensitive to the details of the residual NN force.  A similar analysis is
made for the use of different bound-state single-nucleon wave function
Woods-Saxon (WS) ~\cite{ws} and Hartree-Fock (HF), in the bottom panel.  Here again,  
significant differences up to 20\% arise at higher $T_{\mu}$. 
Similar effects arise for calculations at other incoming energies. The strongest sensitivity,
both for the shape and the magnitude of the cross section, to the
nuclear-structure input occurs at the high-$T_{\mu}$ edges (corresponding to low nuclear excitation energies) of the computed
cross sections.  Concluding, even within the same approach, there is some sensitivity of the cross 
sections to the nuclear-structure input.  We would like to stress that the parametrizations used in our 
calculations are not tuned in any way to the MiniBooNE data. 

\begin{figure*}

\begin{minipage}{\dimexpr\linewidth-0.50cm\relax}
\rotatebox{90}{\mbox{\large \bm{$\textlangle d^{2}\sigma/dT_{\mu}dcos\theta_{\mu} \textrangle (10^{-42}cm^{2}MeV^{-1})$}}}
\vspace*{-8.4cm}\hspace*{16.8cm}
\end{minipage}

\begin{minipage}{\dimexpr\linewidth-0.50cm\relax}
\includegraphics[width=0.94\textwidth]{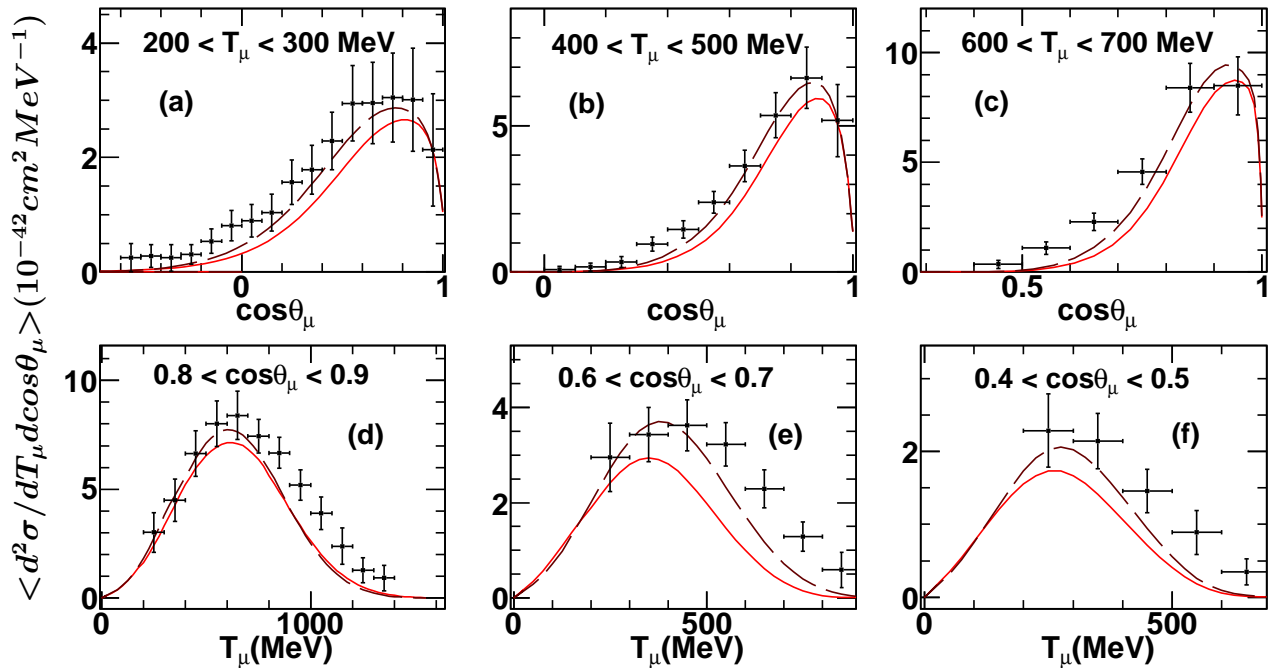}
\caption{(Color online) MiniBooNE flux-folded double-differential
  cross section per target proton for $^{12}$C$(\bar{\nu}_{\mu},
  \mu^{+})X$.  Full (dashed) curves are CRPA with $M_{A} = 1.03$ GeV
  ($M_{A} = 1.35$ GeV). MiniBooNE data are filled squares shown with
  shape uncertainty which excludes an additional 17.2\% normalization
  uncertainty. The top panels show the $\cos\theta_{\mu}$ dependence
  for different ranges of $T_{\mu}$, while the lower panels give the
  $T_{\mu}$ dependence for different ranges of $\cos\theta_{\mu}$.}
  \label{fig8}
\end{minipage}
\end{figure*}

The flux-integrated double-differential cross section for CCQE
antineutrino-nucleus scattering, in terms of the measured quantities
$T_{\mu}$ and $\cos{\theta}_{\mu}$ (hence free from the energy
reconstruction issue ~\cite{martini4,nieves3,lalakulich2012}) can be written as
\begin{eqnarray}
 \left\langle\frac{d^2\sigma}{dT_{\mu}d\cos{\theta}_{\mu}}\right\rangle & = & \frac{1}{\int\Phi(E_{\bar{\nu}})dE_{\bar{\nu}}} \nonumber \\
                                               &   & \int  \left[\frac{d^{2}\sigma}{dT_{\mu}d\cos{\theta}_{\mu}} \right]_{E_{\bar{\nu}}} \Phi(E_{\bar{\nu}})dE_{\bar{\nu}}~, 
\end{eqnarray}
where the antineutrino flux $\Phi(E_{\bar{\nu}})$ is taken
from~\cite{minibooneantinu}. The energy distribution of the MiniBooNE
normalized antineutrino and neutrino flux is shown in
Fig.~\ref{fig2}. The neutrino flux peaks at higher energies than
the antineutrino one.  

In this work, incoming antineutrino energies up to $E_{\overline{\nu}_{\mu}}$ = 2 GeV 
and multipoles up to $J = 12$, are included in the calculations. We have checked that under all
considered kinematic conditions, the computed inclusive antineutrino cross sections do not 
receive sizable contributions from $J > 12$ multipoles. Unless specified otherwise, the used bound-state
single-particle wave functions are solutions to the Schr\"{o}dinger equation with a WS potential.

The double-differential $^{12}$C$(\bar{\nu}_{\mu}, \mu^{+})X$ cross
sections per target proton are displayed in Fig.~\ref{fig3}. The CRPA
and MF calculations are folded with the MiniBooNE $\bar{\nu}_{\mu}$
flux of Fig.~\ref{fig2}. In the dipole axial form factor, we adopt
$M_A=1.03$~GeV which is essentially tuned to deuterium bubble chamber
data. The uncertainties (both with regard to shape and to
normalization) in the MiniBooNE data are not shown.  Comparing the
CRPA and MF results in Fig.~\ref{fig3}, it is clear that the inclusion
of RPA correlations reduces the cross
sections, at the same time shifting the strength towards lower muon
energies. Obviously, both the MF and CRPA calculations reproduce the
major features of the measured $\left( \cos \theta _{\mu} , T_{\mu}
\right)$ distributions: the largest cross sections are for forward
$\theta _{\mu}$ and the peaks shift to smaller $T_{\mu}$ with
increasing $\theta _{\mu}$. This is completely in line with the
expectations from the $\left( \cos \theta _{\mu} , T_{\mu} \right)$
dependence of the $x_B$ and minimum $p_{mis}$ of Fig.~\ref{fig1}.

Figure~\ref{fig4} shows a more detailed picture, displaying double-differential cross
sections as a function of $T_{\mu}$ ($\cos\theta_{\mu}$) for various
bins in the other kinematic variable.  The theoretical results are
obtained by integrating the calculations over the corresponding bin
width. The MiniBooNE data of Fig.~\ref{fig4}  include
the experimental uncertainties. Overall, the CRPA predictions are in satisfactory
agreement with the data. The quality of agreement between the CRPA
calculations and data is best at low and average muon kinetic energies and
forward muon angles. At backward 
$\cos\theta_{\mu}$, the CRPA tends to underestimate the data for higher $T_{\mu}$ values. 
It has been suggested by several authors   
that multinucleon excitations are at the origin of the missing
strength at higher  $T_{\mu}$ and backward $\theta_{\mu}$, as that 
region corresponds with large values of $x_B$ and minimum $p_{mis}$. 
The
quenching due to RPA correlations is strongest at backward $\theta
_{\mu}$ and disappears at the $T_{\mu}$ edges of the
distributions. In general the MF provides a better description of the
data than CRPA both for the shape and     magnitude of the cross section.
A similar observation was made in Ref.~\cite{amaro}, where two
approaches are considered to compute the CCQE ${\nu}_{\mu} + ^{12}$C
cross sections, superscaling and the relativistic mean-field (RMF) approach. 
Of these two,  the RMF model was observed to provide the best description of the shape of the
double-differential cross sections.  Our calculations are in line with those of the RMF
model of \cite{ivanov}, yet slightly closer to the data.

\begin{figure}
\includegraphics[width=0.8\columnwidth]{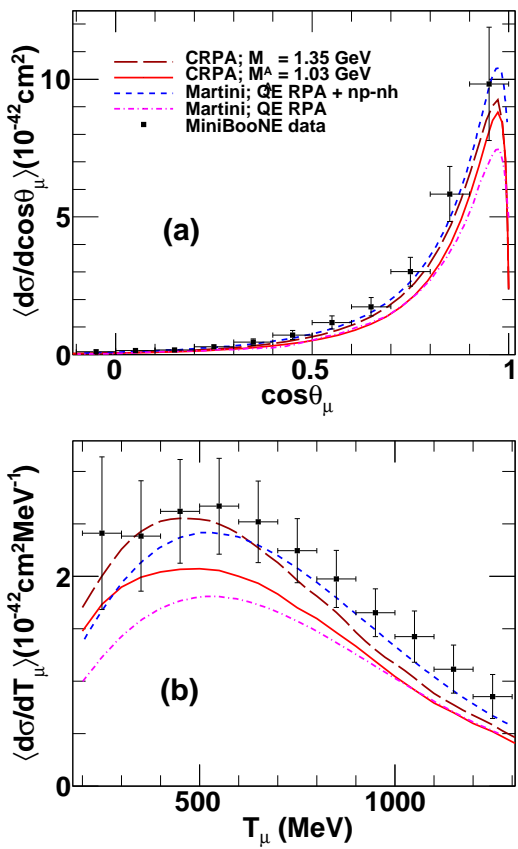}
\caption{(Color online) MiniBooNE flux-folded cross section per target
  proton for $^{12}$C$(\bar{\nu}_{\mu}, \mu^{+})X$ as a function of
  $\cos\theta_{\mu}$ (a) and of $T_{\mu}$ (b). A comparison is
  made of the CRPA cross sections with those of Ref.~\cite{martini}
  (Martini \textit{et al.}). The MiniBooNE data are integrated over $T_{\mu}$
  (a) and over $\cos\theta_{\mu}$ (b)}.
\label{fig10} 
\end{figure}

Various studies have observed different contributions of RPA  and multinucleon effects 
 for  neutrino and antineutrino cross sections. 
The top panel of Fig.~\ref{fig6} shows QE neutrino and
antineutrino cross sections, both normalized to one. In absolute
numbers, the neutrino cross section is always larger, but the
normalized cross section shows that  antineutrino processes 
exhibit a stronger sensitivity to contributions
stemming from the high end of the $T_{\mu}$ spectrum.  As
illustrated in the bottom panel of Fig.~\ref{fig6}, this difference can be explained by
the sign of the transverse interference term in
Eq.~(\ref{sigmaT}). For neutrinos, both transverse terms add
constructively, while for antineutrinos they add
destructively. The absolute value of the interference
contribution to $\sigma_{T}$ is relatively small. Still, for low $T_{\mu}$, the comparable size 
of both transverse  contributions results in a sizable 
 gain of importance of the transverse interference term.
Therefore, at low $T_{\mu}$,  the difference between the $\nu
_{\mu}$ and $\bar{\nu} _{\mu}$ cross sections increases and the antineutrino ones become 
very small.  Hence, the main contribution to antineutrino scattering comes from reactions 
at higher $T_{\mu}$ values and 
antineutrino-nucleus reactions are relatively more sensitive to
low-energy nuclear dynamics than their neutrino counterparts.  
As can be appreciated from Fig.~\ref{fig6}, low nuclear excitation energies
represent a large share of the folded cross sections.  Accordingly,
one may expect that the effect of the RPA correlations is stronger for
$\bar{\nu} _{\mu}A$ interactions.

As a consequence of these differences and the 
respective energy dependence of cross sections, one can also expect differing influences 
of multinucleon effects on neutrino and antineutrino cross sections.
The effect of multinucleon contributions to the $\bar{\nu}_{\mu}$
double-differential cross sections is studied among others in Refs.~\cite{nieves,martini,Amaro:2011aa} and to the
${\nu}_{\mu}$ cross sections in Ref.~\cite{amaro,martini2,Nieves:2011yp}.
From those studies, particularly from Figs.~1 and 4 in Ref.~\cite{amaro},
it emerges that for the very forward-peaked neutrino scattering in  MiniBooNE kinematics, 
multinucleon contributions are responsible for a
significant fraction of the strength at low $T_{\mu}$ and are
essential for reproducing the data. At backward $\theta _{\mu}$, where cross sections 
are very small anyway, the
effect of the multinucleon contributions is rather modest. This can be
understood by realizing that backward $\theta_{\mu}$ corresponds with
larger values of $Q^{2}$ (Fig.~\ref{fig1}). With increasing values of
the range parameter $Q^2$, multinucleon effects naturally lose in
importance \cite{Jan1999}.
 In the superscaling approach of Ref.~\cite{amaro}, it is
argued that the relative impact of \textit{n}p-\textit{n}h contributions increases with
growing  energies of the incoming lepton. 
Moreover, pion-less intermediate
$\Delta$ creation is a source of strength beyond the IA that gains in
importance as one approaches the pole of the $\Delta$ propagator
\cite{Jan1999,Jan2001}. From Fig.~\ref{fig6} it became obvious that the
antineutrino-nucleus reaction has an enhanced sensitivity to the
strength stemming from lower nuclear excitation energies. More specifically in the  MiniBooNE experiment, the antineutrino flux peaks at
lower energies than the neutrino one as shown in Fig.~\ref{fig2}. Under
those kinematic circumstances, one might expect strong  nuclear
effects but reduced \textit{n}p-\textit{n}h contributions through pionless $\Delta$ decay, for example.  

Obviously, modeling the
multitude of np-nh effects at various energies introduces 
uncertainties. Figure~\ref{fig11} shows the predicted
contribution from \textit{n}p-\textit{n}h to the $^{12}$C$(\bar{\nu}_{\mu}, \mu^{+})X$
cross section for two models available in the literature. Whereas the shape of the energy-dependence of the
multinucleon contribution is predicted slightly differently in these studies, its magnitudes differs considerably. 
In both studies, the shape of flux-averaged \textit{n}p-\textit{n}h contributions is similar to that of the QE cross section. 
The divergent views about the role of the \textit{n}p-\textit{n}h illustrate that the model dependencies are
unavoidable given the extensive range of $x_{B}$, $p_{mis}^{min}$, $Q^{2}$ (Fig.~\ref{fig1}) values
covered in the experiments.
 The good general agreement of the calculations is mainly
obtained by the combination of QE and multinucleon contributions,
averaging out the most apparent discrepancies.

The (anti)neutrino-nucleus response calculations require input with
regard to the two vector 
and the axial form factors. They are often parametrized as a
dipole function of the range parameter $Q^{2}$. As a result, each form
factor introduces at least two parameters, a cutoff mass, formally playing the role
of a size parameter and the value at $Q^{2}=0$ that determines the
coupling strength. The two vector form factors are well known from
electron-scattering studies~\cite{escat} and we use a standard dipole
parametrization which is a good approximation for the $Q^{2}$ values probed 
in MiniBooNE (Fig.~\ref{fig1}). The axial form factor, in the dipole form, reads as
\begin{equation}
  G_{A}~=~\frac{g_A}{\left(1+\frac{Q^2}{M_A^{2}}\right)^2}~,\label{ga}
\end{equation}
where $g_A$ is determined from nuclear $\beta$ decay~\cite{betadecay}.
The value $M_A = 1.03\pm0.02$ GeV is regarded as the world's average
value \cite{ma,maworldave1, maworldave2} emerging from bubble-chamber
experiments. Tuning Eq.~(\ref{ga}) to the shape of the $Q^2$ distribution
of the MiniBooNE $\nu_{\mu}$
data~\cite{minibooneantinu,miniboonenucc} favors the value
$M_{A} = 1.35\pm0.17$~GeV.  In Fig.~\ref{fig7}, we investigate the
sensitivity of the computed CRPA cross sections to the adopted value
of $M_{A}$. Changing $M_{A}$ from 1.03 to 1.35~GeV, increases
the cross sections by nearly 10\%. Note that in Fig.~\ref{fig7} we
present the normalized cross sections.  From the figure, it can be
appreciated that modification of $M_{A}$ affects both the energy 
distribution and the $\sigma _{CL} / \sigma _{T}$ ratio. Whereas the overall 
effect of enhancing $M_A$ is a cross section increase, this figure shows that 
more subtle mechanisms are at play.  Enhancing $M_A$ shifts the strength 
to higher nuclear excitation energies, resulting in  a larger impact on the MiniBooNE 
neutrino than antineutrino cross sections.

In Fig.~\ref{fig8}, we study the sensitivity of the double-differential flux-folded CRPA cross section to the 
adopted value of $M_A$.  It can be appreciated that enhancing $M_{A}$ improves the
overall agreement between the CRPA antineutrino cross sections and the
data.  The enhancement is most pronounced at backward muon scattering but still 
does not suffice to bring calculations in agreement with data, especially for 
higher $T_{\mu}$.   
As becomes clear from Fig.~\ref{fig10}, with $M_{A}=1.35$~GeV the
 CRPA results reproduce the data for $T_{\mu}
\le$~600~MeV well.  Under those kinematic conditions, the calculations of
Ref.~\cite{martini} tend to underestimate the data. At higher values of
$T_{\mu}$ the opposite situation occurs with CRPA underestimating the
data.  From the comparison in Fig.~\ref{fig10}, we also find that 
our CRPA cross sections are larger than the QE RPA predictions from
Ref.~\cite{martini}.

The analysis of the MINER$\nu$A antineutrino results \cite{minervanu} favors the transverse enhancement model (TEM).  
In TEM, the magnetic form factors of the bound nucleons are modified in order to
account for the enhancement relative to IA predictions, observed in
the transverse parts of the electron-nucleus cross sections
~\cite{tem}. We stress that in the analysis of Ref.~\cite{minervanu}, the TEM and  $M_{A} =
1.35$~GeV  models predict comparable cross sections at $Q^2 \lesssim
$1~GeV$^2$. Accordingly, one can anticipate that for the $Q^2$ region
accessible at MiniBooNE energies (Fig.~\ref{fig1}), it is difficult to
discriminate between the two effective ways of enhancing the computed weak
responses.


\section{Conclusions}\label{conc}

We have calculated the MiniBooNE flux-folded QE contribution to the
$^{12}$C-antineutrino cross sections and present the results in terms
of the experimentally measured quantities $T_{\mu}$ and
$\cos\theta_{\mu}$. The predictions are made within a nonrelativistic
CRPA. The overall agreement between our predictions for the QE contribution to antineutrino scattering cross sections and the MiniBooNE
measurements is satisfactory. The best description is reached for lower
$T_{\mu}$. 
At higher muon kinetic energies and backward
scattering angles, the CRPA results underestimate the data. 
At larger $T_{\mu}$ one observes a significant sensitivity to the choices made with
regard to the nucleon-nucleon interaction and the single-particle wave-functions. We observe 
that the mean-field cross sections in our calculations are in line with the results of \cite{ivanov} and larger than
those of Fermi-gas calculations. 

As antineutrino cross sections are more sensitive to low-energy nuclear dynamics, 
an effect that becomes even more pronounced owing to energy distribution of the 
MiniBooNE antineutrino flux, the effect of RPA correlations is stronger for  
antineutrinos than for neutrinos. For the MiniBooNE kinematic regime and 
the very forward scattering dominated neutrino interactions, multinucleon mechanism 
can be expected to be most important for reactions with a low-energy outgoing lepton.
Enhancing $M_{A}$ enhances the cross sections 
mostly at higher $T_{\mu}$ and backward scattering angles. Altering $M_A$ has a larger 
influence on neutrino than on antineutrino cross sections.
Still, we observe that in case of antineutrino scattering at MiniBooNE
energies, an enhancement in the nucleon axial mass seems to be
an effective way of improving the quality of agreement between the CRPA calculations and the data,
not only for the size but also for the shape of the double-differential cross section.


\acknowledgments
This research was funded by the Interuniversity Attraction Poles Programme initiated by the Belgian 
Science Policy Office, the Erasmus Mundus External Cooperations Window's Eurindia Project and 
the Research Foundation Flanders (FWO-Flanders).

\end{document}